\documentclass[letterpaper, 10 pt, conference]{ieeeconf}
\IEEEoverridecommandlockouts
\usepackage{xcolor}
\usepackage{cite}
\usepackage{amsmath,amssymb,amsfonts}
\usepackage[english]{babel}
\usepackage{algorithmic}
\usepackage[ruled,vlined]{algorithm2e}
\usepackage{graphicx}
\usepackage{optidef}
\usepackage{listings}
\usepackage{svg}
\usepackage{soul}
\usepackage{pifont}
\usepackage{dirtytalk}
\usepackage{caption}
\usepackage{subcaption}
\usepackage{placeins}
\usepackage{mathtools}
\usepackage{lipsum}
\usepackage{braket}
\usepackage{bbding}
\usepackage{wasysym}
\usepackage{multirow}
\usepackage[export]{adjustbox}
\usepackage{stmaryrd}

\newcommand{\Rspace}{\mathbb{R}}

\newcommand{\Z}{\mathcal{Z}}

\newcommand{\F}{\mathcal{F}}

\newcommand{\fd}[1]{\mathbb{#1}}
\newcommand{\dg}[1]{\mathcal{#1}}

\renewcommand{\int}[2]{\llbracket #1,\ #2 \rrbracket}

\title{Automated Functional Decomposition for Hybrid Zonotope Over-approximations with Application to LSTM Networks}
\author{Jonah J. Glunt, Jacob A. Siefert, Andrew F. Thompson, Justin Ruths, and Herschel C. Pangborn%
\thanks{Jonah J. Glunt, Jacob A. Siefert, Andrew F. Thompson, and Herschel C. Pangborn are with the Department of Mechanical Engineering, The Pennsylvania State University, University Park, PA 16802 USA (e-mail: jglunt@psu.edu; jas7031@psu.edu; thompson@psu.edu; hcpangborn@psu.edu).}
\thanks{Justin Ruths is with the Department of Mechanical Engineering, The University of Texas at Dallas, Richardson, TX 75080 USA (e-mail: jruths@utdallas.edu).}
\thanks{This work was supported by the Department of Defense through the National Defense Science \& Engineering Graduate (NDSEG) Fellowship Program.}
}
\begin{document}

\maketitle
\thispagestyle{empty} 

\begin{abstract}
    Functional decomposition is a powerful tool for systems analysis because it can reduce a function of arbitrary input dimensions to the sum and superposition of functions of a single variable, thereby mitigating (or potentially avoiding) the exponential scaling often associated with analyses over high-dimensional spaces. 
This paper presents automated methods for constructing functional decompositions used to form set-based over-approximations of nonlinear functions, with particular focus on the hybrid zonotope set representation.
To demonstrate these methods, we construct a hybrid zonotope set that over-approximates the input-output graph of a long short-term memory neural network, and use functional decomposition to represent a discrete hybrid automaton via a hybrid zonotope.

\end{abstract}

\section{Introduction}
\label{sect:introduction}
Functional decomposition can be leveraged to efficiently construct a set that over-approximates the graph of a nonlinear function, a tool useful for the verification of nonlinear and hybrid systems~\cite{Siefert_Reach_FuncDecomp}. However, multiple valid decompositions often exist, with some leading to less complexity in the respective set-based analysis.

The Kolmogorov-Arnold Representation Theorem proves that every continuous function which maps from a closed, arbitrary-dimensional hypercube to the real number line can be written as the composition and addition of continuous functions of a single variable~\cite{Kolmogorov1957}. This theorem, and another well-known variant~\cite{Lorentz1962}, have led to advances in approximation theory~\cite{Diaconis1984,Koppen_KolmogorovDecomp}, image processing~\cite{barliga2005image, Leni_KolmogorovImage}, and neural networks~\cite{HechtNielsen1987KolmogorovsMN,Hornik1989_UniversalApproximator, Cybenko1989_UniversalApproximator, liu2024kankolmogorovarnoldnetworks}. 

Work to approximate nonlinear optimization programs \cite{wanufelle2007global} uses functional decomposition to build approximations and to reduce complexity when the same nonlinear component is present in multiple dimensions. Methods in \cite{leyffer2008branch} represent the functional decomposition as a graph network, analyze how to propagate intervals through compositions of functions, and utilize those results to present new techniques for generating Special Ordered Set (SOS) approximations through decomposition. 

The Shunting Yard Algorithm (SYA)~\cite{Dijkstra1961SYA} was developed to parse mathematical expressions by reordering expressions in infix notation into Reverse-Polish Notation (RPN), in which unary and binary operations are placed after their operands, and is still used by many calculators and computers today to parse user inputs into computer language. In~\cite{KVASNICA_2011_AutoDecomp}, the authors represent functional decompositions via a parsing tree and describe an algorithm to generate parsing trees. Additionally,~\cite{KVASNICA_2011_AutoDecomp,SZUCS_2012_OptimalPWA} define \emph{separable} functions as those that can be expressed as the sum of functions of a single variable and show how such functions are easily decomposed into unary functions. Methods for approximating an arbitrary function by constructing a Kolmogorov-like decomposition of a single function are provided in~\cite{Koppen_KolmogorovDecomp}. 

Piecewise affine approximations provide fundamental tools for control and reachability of nonlinear systems~\cite{Bemporad2005,Asarin2003}. Previous work by the authors has 1) shown how a given functional decomposition can be used to generate piecewise affine over-approximations of some specific functions and classes of systems~\cite{Siefert_Reach_FuncDecomp, Siefertthesis_2024}, 2) examined how the over-approximation error of each step in the decomposition is propagated~\cite{Glunt2024_ADHS}, and 3) detailed how such an approximation can be exactly represented as a hybrid zonotope for reachability analysis of nonlinear systems~\cite{Siefertthesis_2024}. Additionally,~\cite{Siefert_Reach_FuncDecomp} analyzed the complexity of constructing a hybrid zonotope for a given functional decomposition, and found that this framework was capable of solving reachability problems that other state-of-the-art methods could not. However, these previous works largely assumed that a decomposition was readily apparent and did not provide detailed procedures for constructing decompositions for arbitrary functions.

\textit{Contributions:} This paper provides novel algorithms to automate functional decomposition of functions of arbitrary dimension. In addition, we provide methods to avoid redundancy and remove excessive decomposition of unary functions. These algorithms are tailored specifically to creating hybrid zonotope over-approximations of nonlinear functions. To the authors' knowledge, this paper is the first to over-approximate a long short-term memory neural network with a hybrid zonotope.

\textit{Outline:} The remainder of this paper is organized as follows. In Section~\ref{sec:Prelims}, we present preliminary materials, including the decomposition of scalar-valued functions. In Section~\ref{sect:AFD}, we propose a framework for automated functional decomposition, including methods for detecting excessive decompositions and handling vector-valued functions. In Section~\ref{sec:examples}, we provide two examples which leverage these algorithms in the context of systems and controls. Concluding remarks are given in Section~\ref{sect:conclusion}.

\section{Preliminaries}
\label{sec:Prelims}
\subsection{Notation}
Matrices are denoted by uppercase letters, e.g., $G\in\Rspace^{n\times n_g}$, and sets by uppercase calligraphic letters, e.g., $\mathcal{Z}\subset\Rspace^{n}$. 
Vectors and scalars are denoted by lowercase letters. 
The $i^{th}$ column of a matrix $G$ is denoted $G_{(\cdot,i)}$.
The $n$-dimensional unit hypercube is denoted by $\mathcal{B}_{\infty}^n=\left\{x\in\Rspace^{n}~|~\|x\|_{\infty}\leq1\right\}$.
The set of all $n$-dimensional binary vectors is denoted by $\{-1,1\}^{n}$ and the closed interval set between a lower bound $b_{l}$ and an upper bound $b_{u}$ is denoted by $\int{b_l}{b_u}$.
Matrices of all $0$ and $1$ elements are denoted by $\mathbf{0}$ and $\mathbf{1}$, respectively, of appropriate dimension.
The concatenation of two column vectors to a single column vector is notated in-line as $(g_1,\ g_2)=[g_1^T\:g_2^T]^T$.
Curly brackets, in addition to denoting sets, are also used to denote optional arguments of a function, e.g., $f(x\{,y\})$ denotes a function of $x$ with an optional argument $y$. Optional arguments begin with a comma just inside of the opening curly bracket to distinguish the difference with set notation. 
When a normally scalar function is passed a vector input, it is implied that the function is to operate element-wise, i.e., if $f:\Rspace\rightarrow\Rspace$ and $x\in\Rspace^n$, then $f(x) \coloneqq (f(x_1), \dots, f(x_n))$.

The set $\mathcal{Z}_h\subset\Rspace^n$ is a \emph{hybrid zonotope} \cite[Def. 3]{Bird_HybZono} if there exists $G^c\in\Rspace^{n\times n_{g}}$, $G^b\in\Rspace^{n\times n_{b}}$, $c\in\Rspace^{n}$, $A^c\in\Rspace^{n_{c}\times n_{g}}$, $A^b\in\Rspace^{n_{c}\times n_{b}}$, and $b\in\Rspace^{n_c}$ such that {\small
    \begin{equation}\label{def-eqn-hybridZono}
        \mathcal{Z}_h = \left\{ \left[G^c \: G^b\right]\left[\begin{smallmatrix}\xi^c \\ \xi^b \end{smallmatrix}\right]  + c\: \middle| \begin{matrix} \left[\begin{smallmatrix}\xi^c \\ \xi^b \end{smallmatrix}\right]\in \mathcal{B}_\infty^{n_{g}} \times \{-1,1\}^{n_{b}}, \\ \left[A^c \: A^b\right]\left[\begin{smallmatrix}\xi^c \\ \xi^b \end{smallmatrix}\right] = b \end{matrix} \right\}\:.
\end{equation}}%
A hybrid zonotope is the union of $2^{n_b}$ constrained zonotopes (polytopes) corresponding to the possible combinations of binary factors $\xi^b$. 

\subsection{Functional Decomposition}
\label{sec:funcDecomp}
Explicit functions can be decomposed into unary (scalar input, scalar output) and binary (two scalar inputs, one scalar output) functions. Approximating unary and binary functions avoids the exponential memory complexity of piecewise approximations with respect to the argument dimension \cite{leyffer2008branch}. A function $f:\Rspace^n\rightarrow\Rspace^m$ is decomposed by introducing intermediate variables
\begin{align}
    \label{eqn-decomp}
    w_j & =
    \begin{cases}
    x_j\ , & j\in\{1,\dots,n\}\;,\\
    h_j(w_{i}\{,w_{k}\})\ , & j\in\{n+1,\dots,n+K\}\;,
    \end{cases}
\end{align}
where $i,k<j$, giving $f(x) = (w_{n-K-m+1},\dots,w_{n+K})$. We refer to the intermediate decomposition variables $w_j$ as \emph{observables}.
The first $n$ assignments directly correspond to the $n$ elements of the argument vector $x$, assignments $n+1,\dots,n+K$ are defined by the unary function or binary functions $h_j$, and the final $m$ assignments are associated with $f(x)$. In the case that $h_j$ is unary, the second argument is omitted. Many examples of functional decompositions developed manually are given in~\cite{Siefert_Reach_FuncDecomp,Siefertthesis_2024}. This paper focuses on developing an automated algorithmic method for decomposing functions. 

\subsection{Graphs of Functions}
\label{sec:Intro_GOFs}
Given a mapping $\phi:D_\phi \rightarrow \mathcal{Q}$, we refer to the set $\Gamma(\phi) = \{(p,\phi(p))\mid p\in D_\phi\}\subset D_\phi\times\mathcal{Q}$ as the \emph{graph of the function $\phi$}. The set $\text{D}_{\phi}$ is referred to as the \emph{domain set} of $\phi$ and can be chosen by a user as the set of inputs of interest. A set $\Z\subset D_\phi \times \mathcal{Q}$ is said to \emph{over-approximate} $\Gamma(\phi)$ if $(p,\ \phi(p))\in\Z\;\forall p\in D_\phi$.
Procedures for constructing exact and over-approximated graphs of functions from a given functional decomposition are given in~\cite[Thm. 4 and Cor. 1-4]{Siefert_Reach_FuncDecomp}.

\subsection{Graphs (Distinct from Graphs of Functions)}
\label{sec:Intro_Graphs}

A graph $\mathcal{G}$ consists of a set of $n_V$ vertices $V=\{V_1, V_2, \dots, V_{n_V}\}$ and a set of $n_E$ edges $E$ which connect pairs of vertices. The edge connections of $\mathcal{G}$ can be represented as an adjacency matrix $A\in \Rspace^{n_V \times n_V }$ where
\begin{equation}
    A_{i,j} = 
    \begin{cases}
        1\ , & \text{if } V_i \text{ is connected to } V_j\;,\\
        0\ , & \text{else.}
    \end{cases}%
\end{equation}%
If $\mathcal{G}$ is an undirected graph, then $A$ is a symmetric matrix, i.e., $A_{i,j} = A_{j,i}$. However, if $\mathcal{G}$ is a directed graph, as is the assumption in the remainder of this paper, then this is not necessarily the case. If there exists an edge with vertices $V_i$ as its tail and $V_j$ as its head (i.e., $A_{i,j} = 1$), then $V_j$ is a successor of $V_i$ and $V_i$ is a predecessor of $V_j$. The transpose of a graph, $\mathcal{G}'$, is a directed graph with the same vertices as $\mathcal{G}$, but with edges of opposite orientation (i.e., the adjacency matrix of $\mathcal{G}'$ is $A^T$). For a vertex $V_i$, the number of edges entering it, referred to as its indegree, is denoted by $d^-(V_i)$, and the number of edges leaving the vertex, referred to as its outdegree, is denoted by $d^+(V_i)$. A \emph{walk} in graph $\mathcal{G}$ is an ordered list of connected vertices, and a \emph{reverse walk} is a walk in the transposed graph, $\mathcal{G}'$. A directed graph is said to be acyclic if $\forall V_i \in V$, there exists no possible walk from $V_i$ to itself.

\section{Automated Functional Decomposition}
\label{sect:AFD}

This section presents methods to automate the production of a functional decomposition in the form of \eqref{eqn-decomp} from an expression given in infix notation, which represents algebraic functions using mathematical operators placed between operands. The process uses an existing algorithm to convert from infix notation to another standard notation and then presents a novel algorithm to convert the result to a functional decomposition. The proposed approach addresses two challenges when constructing functional decompositions for use in set-theoretic analyses. Namely, the removal of redundant observables and excessive decomposition of unary functions.

\subsection{Reverse Polish Notation}
\label{sect:RPN_SYA}
Infix notation is commonly used to represent mathematical expressions with operators placed between operands, e.g., $x+y*z$. Reverse Polish Notation (RPN), also referred to as reverse Likasiewicz notation or postfix notation, represents a mathematical expression with operands preceding their operators. The Shunting Yard Algorithm (SYA)~\cite{Dijkstra1961SYA} uses precedence and associativity to parse infix notation and outputs an equivalent expression in RPN, e.g., 
\begin{align}
\label{eqn-RPNex1}
    \begin{matrix}
        \text{\underline{Infix Notation}} & \text{\underline{SYA}} & \text{\underline{RPN}} &\\
        x+y\times z & \rightarrow & x\ y\ z\ \times\ + &,\\
        3\times y\times\cos(x)^2 & \rightarrow & 3\ y\ x\ \cos\ 2\ \hat{}\ \times\ \times &.
    \end{matrix}
\end{align}
We refer to each number, variable, operator, and function as a \emph{token}. The top and bottom RPN expressions in \eqref{eqn-RPNex1} have 5 and 8 tokens, respectively. To parse RPN, start at the left of the expression and \emph{push} tokens onto a stack until a operator or function token is reached. Then \emph{pop} the appropriate number of tokens off of the stack, perform the operation to produce a single token, and \emph{push} the resulting token to the stack. The process is repeated until there are no remaining tokens in the expression and the result is on the stack. 

\subsection{RPN $\rightarrow$ Functional Decomposition for Scalar-valued Functions}
\label{sect:RPN2FD_1D}
A string of tokens is denoted by $\mathcal{E}$, and $\mathcal{E}_1$ denotes the first token of $\mathcal{E}$. As tokens are pushed or popped to a stack, the value of $\mathcal{E}_1$ changes. For ease of notation and without loss of generality, we assume the input variables of the expression of interest are $w_1,\dots,w_{n_x}$ where $n_x$ is the number of input variables. \textbf{Algorithm~\ref{alg:RPN2FD}} provides an automated method to convert a string of tokens in RPN into a functional decomposition of the form \eqref{eqn-decomp} by leveraging the order of operations and evaluations of sub-expressions to generate observables. \textbf{Algorithm~\ref{alg:RPN2FD}} has two apparent flaws when used with piecewise affine and piecewise polytopic approximations, which we describe as 1) redundant observables, and 2) excessive decomposition of unary functions.

\begin{algorithm}[t]
    \SetAlgoLined
    \LinesNumbered
    \KwResult{Functional Decomposition}
    $k \leftarrow n_x+1$\\
    \While{$\mathcal{E}$ is not empty}{
        \While{$\mathcal{E}_1$ is a number or a variable}{
            \textbf{push} $\mathcal{E}_1$ to stack.\\
        }
        \textbf{pop} tokens from stack associated with the operator or function $\mathcal{E}_1$, form observable expression and save as $w_k$.\\
        $k \leftarrow k+1$
    }
     \caption{RPN $\rightarrow$ Functional Decomposition $\mathcal{O}(n)$: Given a string of $n$ tokens $\mathcal{E}$ in RPN
     }
     \label{alg:RPN2FD}
\end{algorithm} 

\paragraph{Redundant Observables} Consider the expression $\sin(x) + \sin(x)^2$. The SYA yields the RPN expression
\begin{align}
\label{eqn-ROex-RPN}
    x\ \sin\ x\ \sin\ 2\ \wedge\ + \qquad,   
\end{align}%
and \textbf{Algorithm~\ref{alg:RPN2FD}} gives the functional decomposition
\begin{align}
    \nonumber
    \text{Input}&:\ w_1=x\;,\\
    \nonumber
    w_2 &= \sin(w_1)\:,\\
    \label{eqn-UUDex}
    w_3 &= \sin(w_1)\:,\\
    \nonumber
    w_4 &= w_3^2\:,\\
    \nonumber
    w_5 &= w_3 + w_4\:.
\end{align}
Note that $w_2$ and $w_3$ are redundant as they represent the same quantity. When considering use of the functional decomposition with piecewise affine or piecewise polytopic approximations of nonlinear functions, this redundancy should be avoided because 1) unnecessary complexity will be accrued in approximating the quantity $\sin(w_1)$ twice instead of once, and 2) when creating approximations of these nonlinear functions, $w_2$ and $w_3$ will be uncoupled, i.e., the approximations of $w_2$ and $w_3$ may evaluate to different values given the same $w_1$.

\begin{figure}
    \centering
    \includegraphics[width=0.75\linewidth]{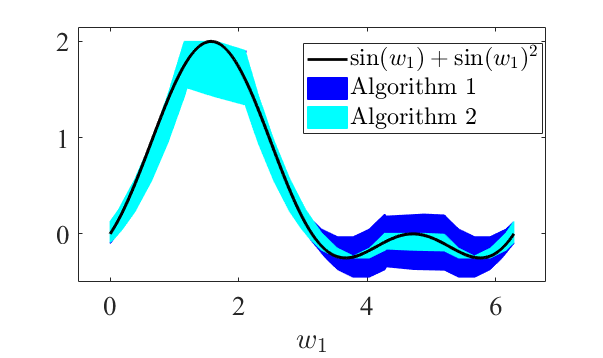}
    \caption{Over-approximations of the graph of  \eqref{eqn-ROex-RPN}, obtained using functional decompositions generated by \textbf{Algorithm~\ref{alg:RPN2FD}} (dark blue and cyan) and \textbf{Algorithm~\ref{alg:RPN2FD_mod}} (cyan only).}
    \label{fig:ExUUD}
    \vspace{-\baselineskip}
\end{figure}

To address this, \textbf{Algorithm~\ref{alg:RPN2FD_mod}} builds on \textbf{Algorithm~\ref{alg:RPN2FD}} to remove redundant observables by checking to see if an equivalent observable already exists. A new observable is only created if it is unique from all previous observables. Applying \textbf{Algorithm~\ref{alg:RPN2FD_mod}} to the RPN expression given in \eqref{eqn-ROex-RPN} results in the more compact functional decomposition%
\begin{align}
    \nonumber
    \text{Input}&:\ w_1=x\;,\\
    \label{eqn-UUDex-Red}
    w_2 &= \sin(w_1)\:,\\
    \nonumber
    w_3 &= w_2^2\:,\\
    \nonumber
    w_4 &= w_2 + w_3\:.
\end{align}%
Figure \ref{fig:ExUUD} exemplifies the benefits of accounting for redundant observables (\textbf{Algorithm~\ref{alg:RPN2FD}} vs. \textbf{Algorithm~\ref{alg:RPN2FD_mod}}). In this case, piecewise polytopic over-approximations of the nonlinear sinusoidal and quadratic observables were made with a tolerance of $\pm 0.1$ and $\pm 0.01$, respectively. Because \textbf{Algorithm~\ref{alg:RPN2FD}} results in a redundant observable, the resulting functional decomposition is both less accurate and must store 8 additional piecewise approximations. 
Checking for redundant observables does incur additional computational cost, as \textbf{Algorithm~\ref{alg:RPN2FD}} scales with the number of tokens $n$ as $\mathcal{O}(n)$ whereas \textbf{Algorithm~\ref{alg:RPN2FD_mod}} scales as $\mathcal{O}(n^2)$. However, this additional computational cost is often worthwhile to obtain a more accurate and compact functional decomposition.

\begin{algorithm}
    \SetAlgoLined
    \LinesNumbered
    \KwResult{Functional Decomposition: $w_1,...,w_{n_x+K}$}
    $k=n_x+1$\\
    \While{$\mathcal{E}$ is not empty}{
        \While{$\mathcal{E}_1$ is a number or a variable}{
            \textbf{push} $\mathcal{E}_1$ to stack.\\ 
        } 
        \textbf{pop} tokens from stack associated with the operator or function $\mathcal{E}_1$ and form \emph{candidate} observable expression, $w_c$. \\
           \eIf{$w_c\neq w_i\ \forall i\in \{1,...,k-1\}$}{
            save $w_k = w_c$\\
            $k \leftarrow k+1$
            }{
            substitute previously defined and equivalent observable for $w_c$ and \textbf{push} to stack.
            } 
    }
     \caption{Modified RPN $\rightarrow$ Functional Decomposition $\mathcal{O}(n^2)$: Given a string of $n$ tokens $\mathcal{E}$ in RPN
     }
     \label{alg:RPN2FD_mod}
    \end{algorithm}

\paragraph{Excessive Decomposition of Unary Functions} Consider the expression $\cos(\sin(x_1\times x_2)) + \sin(\cos(\sin(x_1\times x_2))) +\sin(x_1\times x_2)$.
The SYA yields the RPN expression
\begin{align}
\begin{split}
    &x_1\ x_2\ \times\ \sin\ \cos\ x_1\ x_2\ \times\ \dots\\
    &\sin\ \cos\ \sin\ +\ x_1\ x_2\ \times\ \sin\ + \qquad,
\end{split}
\end{align}
and \textbf{Algorithm~\ref{alg:RPN2FD_mod}} gives the functional decomposition
\begin{align} \label{eq:nested}
\nonumber
    \text{Inputs}&:\ w_1=x_1,\ w_2=x_2 \:,\\
    \begin{split}
    w_3 &= w_1 \times w_2\:,\\
    w_4 &= \sin(w_3)\:,\\
    w_5 &= \cos(w_4)\:,\\
    w_6 &= \sin(w_5)\:,\\
    w_7 &= w_5 + w_6\:,\\
    w_8 &= w_4 + w_7.
    \end{split}
\end{align}%
A more concise decomposition may be realized by combining $w_4$, $w_5$, $w_6$, $w_7$, e.g.,
\begin{align} \label{eq:un-nested}
\nonumber
    \text{Inputs}&:\ w_1=x_1,\ w_2=x_2 \:,\\
    \begin{split}
    w_3 =& w_1 \times w_2\:,\\
    w_8 =& \sin(\cos(\sin(w_3)))+ \cos(\sin(w_3))+\sin(w_3)\:.
    \end{split}
\end{align}%
This more compact functional decomposition allows for the approximation of fewer nonlinear functions. These excessive unary decompositions can be detected and eliminated by representing the functional decomposition structure as a directed acyclic graph. For example, the composition of \eqref{eq:nested} can be represented as a graph with adjacency matrix
\begin{equation}
    \label{eq:nestedAdj}
    A = {\tiny\begin{bmatrix}
    0 & 0 & 0 & 0 & 0 & 0 & 0 & 0\\
    0 & 0 & 0 & 0 & 0 & 0 & 0 & 0\\
    1 & 1 & 0 & 0 & 0 & 0 & 0 & 0\\
    0 & 0 & 1 & 0 & 0 & 0 & 0 & 0\\
    0 & 0 & 0 & 1 & 0 & 0 & 0 & 0\\
    0 & 0 & 0 & 0 & 1 & 0 & 0 & 0\\
    0 & 0 & 0 & 0 & 1 & 1 & 0 & 0\\
    0 & 0 & 0 & 1 & 0 & 0 & 1 & 0\\
    \end{bmatrix}}\:,
\end{equation}
where $A_{i,j}\in\{0,1\}$ and $A_{i,j}=1 \iff w_i$ is an argument of $h_j(\cdot)$. 

Detection of an excessive unary decomposition can be achieved by finding a pair of vertices, $V_i$ and $V_j$, $i\neq j$, such that observable $w_j$ is only a function of $w_i$. Then, the subgraph between $V_i$ and $V_j$ can be replaced by a single edge which captures the unary function composition. The proposed detection process, given as \textbf{Algorithm~\ref{alg:ReduceUUD}}, starts by forming two sets for each vertex. The first set, $\mathcal{W}_i$, is the intersection of travelled vertices on all forward walks from $V_i$ and the second set, $\mathcal{M}_i$, is the intersection of travelled vertices on all reverse walks from $V_i$, i.e.,  $\mathcal{W}_i$ and $\mathcal{M}_i$ contain vertices that \textit{must} be visited on every forward and reverse walk, respectively. The graph is searched for pairs of vertices $V_i$ and $V_j$ which have mutual dependency in the original and transposed graphs, i.e., $V_j \in \mathcal{W}_i$ and $V_i \in \mathcal{M}_j$.

To perform the simplification step of this method, we define a recursive composition function \texttt{comp($\cdot,\cdot$)} as%
\vspace{-6pt}
\begin{align}
    &\texttt{comp}(V_i, V_j) = \nonumber \\
    &\begin{cases}
        w_i\ , & \text{if } i=j\;, \\
        h_j(\texttt{comp}(V_i, \mathcal{P}_{j,1})\{, \texttt{comp}(V_i, \mathcal{P}_{j,2})\}), &\text{else.}
    \end{cases}
\end{align}
where $\mathcal{P}_j$ is the set of one or two predecessors of $V_j$. The result of \texttt{comp($V_i,V_j$)} is the composition of unary and binary functions with the single input $w_i$. A graphical representation of reducing \eqref{eq:nested} to \eqref{eq:un-nested} is demonstrated in Figure~\ref{fig:asgS}. Note that \textbf{Algorithm~\ref{alg:ReduceUUD}} does not simplify the graph in Figures~\ref{fig:asgS}(a)-(c) as $V_1\notin \mathcal{M}_3$, $V_2\notin \mathcal{M}_3$, and $d^+(V_3)=1$, respectively. Repeating \textbf{Algorithm~\ref{alg:ReduceUUD}} until it no longer reduces the amount of observables will result in the desired and concise functional decomposition.

\textbf{Algorithm~\ref{alg:ReduceUUD}} is similar to a process known as \emph{smoothing away} vertices of degree two from a graph~\cite{Gross2018_GraphTheory}; more generally, this is a form of \emph{edge contraction}~\cite{Wolle2004}. While the idea of edge contraction is not novel to this paper, we are not only interested in smoothing out vertices of degree 2, but all vertices that have mutual dependency in both the original and transposed graphs.  

\subsection{RPN $\rightarrow$ Functional Decomposition for Vector-valued Functions}
\label{sect:RPN2FD_MD}

The proposed method in Section~\ref{sect:RPN2FD_1D} completes and simplifies the functional decomposition of a scalar-valued function. We now expand this to be compatible with a vector-valued function $h : \Rspace^{n_x} \rightarrow \Rspace^{n_y}$. Creating a functional decomposition for each element of $h$ may result in redundant observables across decompositions, e.g.,
\begin{equation}
    h(x) =\begin{bmatrix}
        \sin(x)\\ \cos(\sin(x))
    \end{bmatrix}\:,
\end{equation}
has similar terms in $h_1$ and $h_2$. The functional decomposition of each element would be
\begin{align}
    \begin{tabular}{r l r l}
        \text{Input:} & $w_1=x\;,$ & \text{Input:} & $w_1=x\;,$ \\
        $w_2=$ & $\sin(w_1)$\;, & $w_3=$ & $\sin(w_1)\;,$\\
         & & $w_4=$ & $\cos(w_3)\;.$
    \end{tabular}
\end{align}

The simplest way to address this is to concatenate the functional decompositions of each element of $h$ into one single decomposition, ensuring the observables maintain unique indices, e.g.,
\begin{align}
    \begin{tabular}{r l r l}
        \text{Input:} & $w_1=x\;,$ \\
        $w_2=$ & $\sin(w_1)\;,$\\
        $w_3=$ & $\cos(w_2)\;.$\\
    \end{tabular}
\end{align}
One way to accomplish this while protecting output observables from simplification is to introduce a new function $\bullet(\cdot)$, which evaluates to its argument, i.e., $\bullet(x) = x$, and run \textbf{Algorithm~\ref{alg:ReduceUUD}} on the function $\sum_{i = 1}^{n_y} \bullet(h_i)$. 
This prevents redundant observables across elements of $h$ as these are detected using \textbf{Algorithm~\ref{alg:RPN2FD_mod}} and allows for easy identification of output observables of $h$.

While performing the conversion from RPN to a functional decomposition in \textbf{Algorithm~\ref{alg:RPN2FD_mod}}, output observables are identified as the tokens on which the identity operators act. When $\mathcal{E}_1 = \bullet(\cdot)$, we can simply pop one token from the stack, note which observable it corresponds to, push it back onto the stack, and discard $\mathcal{E}_1$. The vertices corresponding to these noted observables will be added to the set of protected vertices.

Again, consider the functional decomposition given in \eqref{eq:nested}, but with two outputs $w_4$ and $w_8$ which are added to a set of protected variables. \textbf{Algorithm~\ref{alg:ReduceUUD}} results in the functional decomposition
\begin{align}
\nonumber
    \text{Inputs}&:\ w_1=x_1\;, w_2=x_2 \:,\\
    \begin{split}
    w_3 &= w_1 \times w_2\:,\\
    w_4 &= \sin(w_3)\:,\\
    w_8 &= \sin(\cos(w_4)) + \cos(w_4) + w_4,
    \end{split}
\end{align}
and the process is depicted in Figure~\ref{fig:asgV}.

\newcommand{\subplotheightgraph}{1.8in}
\newcommand{\subplothorizontalgapgraph}{-1.5cm}
\begin{figure*}[h!]
    \centering
    \begin{subfigure}[b]{\subplotheightgraph}
         \centering
         \includegraphics[height=\subplotheightgraph]{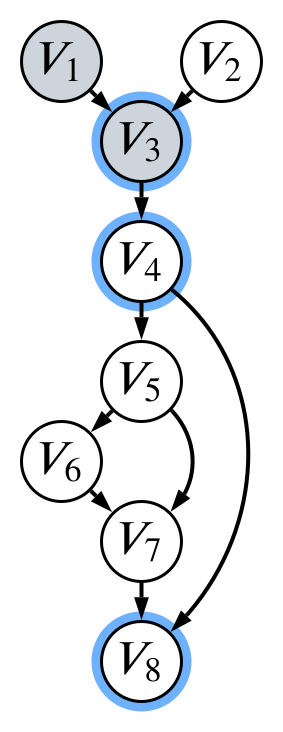}
         \caption{$(i,j) = (1,3)$}
         \label{fig:asgS-a}
    \end{subfigure}\hspace{\subplothorizontalgapgraph}
    \begin{subfigure}[b]{\subplotheightgraph}
         \centering
         \includegraphics[height=\subplotheightgraph]{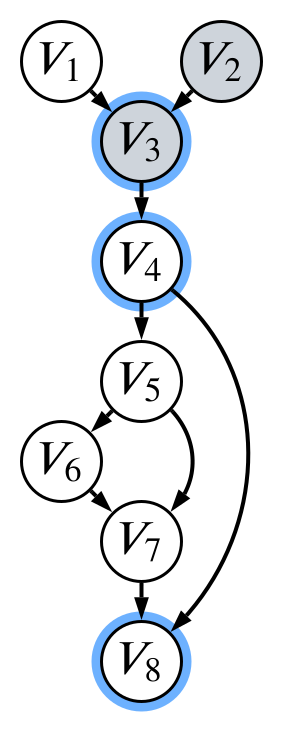}
         \caption{$(i,j) = (2,3)$}
         \label{fig:asgS-b}
    \end{subfigure}\hspace{\subplothorizontalgapgraph}
    \begin{subfigure}[b]{\subplotheightgraph}
         \centering
         \includegraphics[height=\subplotheightgraph]{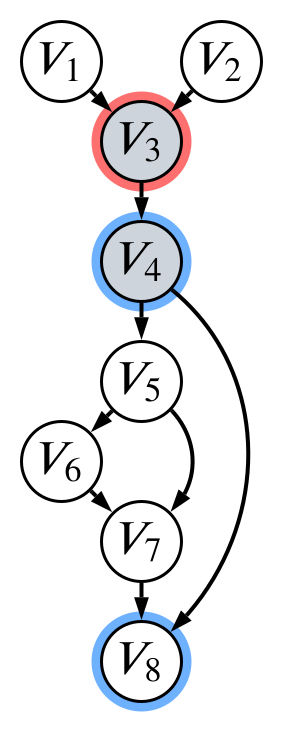}
         \caption{$(i,j) = (3,4)$}
         \label{fig:asgS-c}
    \end{subfigure}\hspace{\subplothorizontalgapgraph}
    \begin{subfigure}[b]{\subplotheightgraph}
         \centering
         \includegraphics[height=\subplotheightgraph]{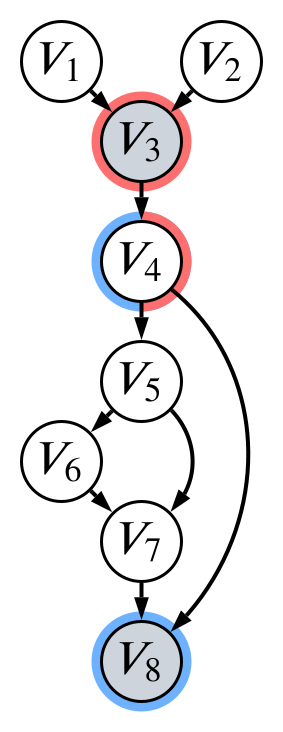}
         \caption{$(i,j) = (3,8)$}
         \label{fig:asgS-d}
    \end{subfigure}\hspace{\subplothorizontalgapgraph}
    \begin{subfigure}[b]{\subplotheightgraph}
         \centering
         \includegraphics[height=\subplotheightgraph]{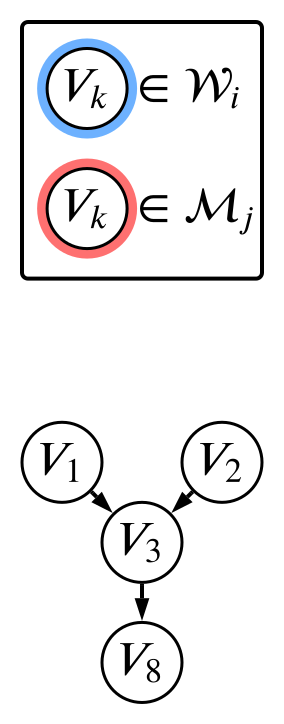}
         \caption{New graph}
         \label{fig:asgS-e}
    \end{subfigure}\hspace{\subplothorizontalgapgraph}
    \caption{Graph representation of~\eqref{eq:nested} being simplified to~\eqref{eq:un-nested} according to \textbf{Algorithm~\ref{alg:ReduceUUD}}. Gray-filled vertices indicate $V_i$ and $V_j$. Blue and red-outlined vertices indicate elements of the intersection of forward walks $\mathcal{W}_i$ from $V_i$ and reverse walks $\mathcal{M}_j$ from $V_j$, respectively.  (a) $V_1 \notin \mathcal{M}_3$. (b) $V_2 \notin \mathcal{M}_3$. (c) $A_{i,j}=0$. (d) This iteration satisfies all conditions in lines 6, 7, and 8 of \textbf{Algorithm~\ref{alg:ReduceUUD}}. (e) The resulting graph from lines 9 and 10 of \textbf{Algorithm~\ref{alg:ReduceUUD}}.}
    \label{fig:asgS}

    \vspace{-0.2in}
\end{figure*}

\begin{figure}[htb!]
    \centering
    \begin{subfigure}[b]{\subplotheightgraph}
         \centering
         \includegraphics[height=\subplotheightgraph]{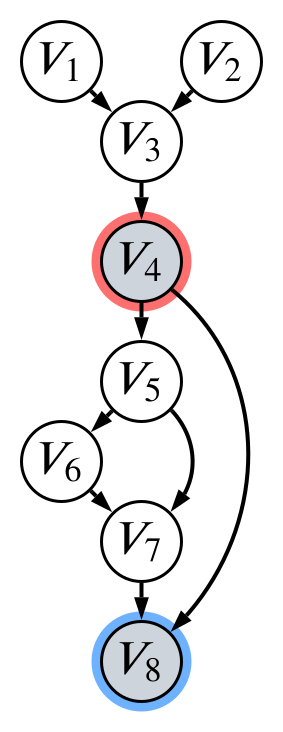}
         \caption{$\begin{aligned}
             i=4\\ j=8
         \end{aligned}$}
         \label{fig:asgV-a}
    \end{subfigure}\hspace{\subplothorizontalgapgraph}
    \begin{subfigure}[b]{\subplotheightgraph}
         \centering
         \includegraphics[height=\subplotheightgraph]{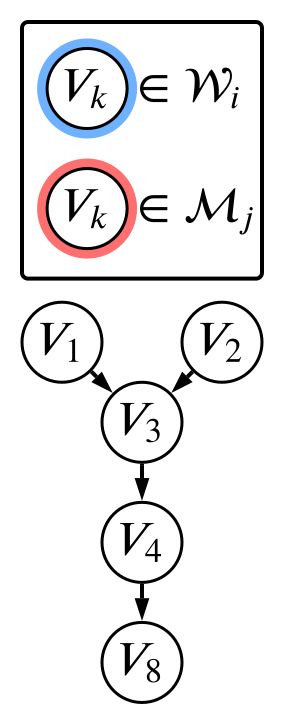}
         \caption{$\begin{aligned}
             \text{New } \\ \text{graph}
         \end{aligned}$}
         \label{fig:asgV-b}
    \end{subfigure}
    \caption{Graph representation of~\eqref{eq:nested} being simplified to~\eqref{eq:un-nested} according to \textbf{Algorithm~\ref{alg:ReduceUUD}} with $V_4$ and $V_8$ listed as protected vertices.}
    \label{fig:asgV}

    \vspace{-0.2in}
\end{figure}

\begin{algorithm}
    \SetAlgoLined
    \LinesNumbered
    \KwResult{$\fd{H}_s$}
    generate directed graph $\dg{G}$ from $\fd{H}$, with vertices $V$, edges $E$, and adjacency matrix $A$.\\
    \For{$V_i\in V$}{generate $\mathcal{W}_i$ and $\mathcal{M}_i$.}
    \For{$i\in\{1,...,n_V\}$}{
    \For{$V_j\in\mathcal{W}_i$}{
    \If{$V_i \in \mathcal{M}_j$}{
    \If{$A_{i,j} = 0$ OR $d^+(V_i)>1$}{
    $w_j \leftarrow \texttt{comp}(V_i, V_j)$\\
    remove all observables $w_k$ associated with vertices visited on all walks between $V_i$ and $V_j$, $k\neq i,j$.    
    }
    }
    }
    }
    $\fd{H}_s \leftarrow$ functional decomposition from $\mathcal{G}$ with re-indexed observables. 
     \caption{Given a functional decomposition $\fd{H}$, generate a functional decomposition $\fd{H}_s$ which has fewer unary redundancies.}
     \label{alg:ReduceUUD}
    \end{algorithm}

\subsection{Permissible Multi-input Functions}
Depending on the context of how a functional decomposition will be used, it may not always be desired to decompose a function into purely unary functions and addition. When a given set representation can exactly represent a given multivariable operation, then there is no reason to further decompose that function. For example, hybrid zonotopes can represent affine mappings exactly, so affine functions of any number of inputs may be considered valid in a decomposition. If a different set representation is used, for example polynomial zonotopes~\cite{SparsePolyZono_Althoff}, then affine mappings and certain polynomial expressions may be useful in a decomposition. This can be accomplished in \textbf{Algorithm~\ref{alg:RPN2FD_mod}} by defining specific operators in line 3 that one wishes to \textbf{push} instead of \textbf{pop}, allowing them to be combined into the same observable expression.
\section{Numerical Examples}
\label{sec:examples}

\subsection{Long Short-term Memory Neural Networks}
\label{sec:LSTM}

Long short-term memory (LSTM) networks are a recurrent neural network (RNN) architecture first developed to store information over extended time intervals (on the order of thousands of discrete steps)~\cite{LSTM}. Where other recurrent architectures previously struggled with this task due to the vanishing gradient effect, LSTM networks introduced gated memory states that are passed between network prediction iterations, and can be opened or closed via multiplication with other special units. An LSTM network with $N$ nodes has two vectors that it computes and passes to the network at the next step: The \emph{hidden state} $h_t\in\int{-1}{1}^N$ and the \emph{cell state} $c_t\in\int{-1}{1}^N$, where the subscript $t$ is used to indicate the $t$-th step. The input to the LSTM network, $x_t\in\Rspace^d$, is combined with the past hidden state to make the next update. The following values are then computed:

\vspace{-11pt}
{\small\begin{subequations}
\label{eqn:LSTM}
\begin{align}
    f_t &= \sigma_g(W_f\cdot(h_{t-1},\ x_t) + b_f)\;, \\
    i_t &= \sigma_g(W_i\cdot(h_{t-1},\ x_t) + b_i)\;,\\
    \tilde{c}_t &= \sigma_c(W_c\cdot(h_{t-1},\ x_t) + b_c)\;,\\
    c_t &= f_t \odot c_{t-1} + i_t \odot \tilde{c}_t\;,\\
    o_t &= \sigma_g(W_o\cdot(h_{t-1},\ x_t) + b_o)\;,\\
    h_t &= o_t \odot \sigma_c(c_t)\;,
\end{align}
\end{subequations}}%
where $\sigma_g:\Rspace^N\rightarrow\int{0}{1}^N$ is the \emph{gate function}, $\sigma_c:\Rspace^N\rightarrow\int{-1}{1}^N$ is the \emph{state function}, and $\odot$ is the Hadamard product (element-wise multiplication) operator. In addition to being stored in memory for the next iteration, $h_t$ is also the output of the LSTM network. The weight matrices $W_f,W_i,W_c,W_o\in\Rspace^{N\times(N+d)}$ and bias vectors $b_f,b_i,b_c,b_o\in\Rspace^N$ are learned during training. Typically, $c_0$ and $h_0$ are initialized to zero, and then evolve with each step of the LSTM network. 

While set-based reachability analysis and verification of feedforward neural networks has been studied extensively in the literature~\cite{Ortiz2023,Zhang2023}, much less attention has been paid to RNN architectures. Until recently, approaches were limited to unrolling the recurrent structure into a feedforward structure as long as the input data \cite{akintunde2019verification}, or over-approximating the RNN output by a feedforward network \cite{jacoby2020verifying}. Star sets have been used to exactly compute reachable sets of ReLU-only RNNs \cite{tran2023verification}, and very recently, sparse star sets have been used for reachability analysis of multiple RNN architectures, including LSTM networks~\cite{Choi2025_RNN_StarSets}. Compared with feedfoward neural networks (especially those with only ReLU activation functions), LSTM networks have significantly more complicated nonlinear architectures within each recurrent step due to the gating operations.

As seen in~\eqref{eqn:LSTM}, there are already at least six equations per hidden node required to define the behavior of this network. By applying \textbf{Algorithm 1}, the LSTM network can be decomposed into the functional decomposition
\newcommand{\inputs}{(h_{t-1},\ x_t)}
\allowdisplaybreaks
\begin{equation} \label{eq:lstm_decomp}{\scriptsize\begin{array}{l | l}
    \text{Inputs}:\ x_t,\ h_{t-1},\ c_{t-1} \:, & \lambda =\theta^2\;,\\
    \alpha = W_f\cdot\inputs+b_f\;, & \mu = 0.25\kappa - 0.25\lambda + \beta\odot c_{t-1}\;,\\
    \beta = \sigma_g(\alpha)\;, & \xi = W_o\cdot\inputs+b_o\;,\\ 
    \gamma = W_i\cdot\inputs+b_i\;, & \rho = \sigma_g(\xi)\;,\\
    \delta = \sigma_g(\gamma)\;, & \tau = \sigma_c(\mu)\;,\\ 
    \varepsilon = W_c\cdot\inputs + b_c\;, & \upsilon = \rho + \tau\;,\\
    \zeta = \sigma_c(\varepsilon)\;, & \varphi = \rho - \tau\;,\\ 
    \eta = \delta + \zeta\;, & \chi = \upsilon^2\;,\\
    \theta = \delta - \zeta\;, & \psi = \varphi^2\;,\\
    \kappa = \eta^2\;, & \omega = 0.25\chi - 0.25\psi \leftarrow\text{Outputs}\;.    
\end{array}}\end{equation}
For the sake of clarity, this decomposition was written using vectored variables $\alpha$, $\beta$, etc. instead of subscripted indices $w_1$, $w_2$, etc., but it still functions in the same manner as~\eqref{eqn-decomp}. The full decomposition via \textbf{Algorithm 1} has over 100 scalar observables, which are not reduced in number by \textbf{Algorithms 2} and \textbf{3} in this case. However, with more than 100 observables to keep track of, performing this decomposition entirely by hand could be prohibitively time consuming even for this relatively small network; thus, the proposed methods serve as an enabling tool by automating the decomposition process. 

In the decomposition~\eqref{eq:lstm_decomp}, multi-input decomposition expressions are permitted when they are affine. This is because hybrid zonotopes can represent affine functions exactly, so no further decomposition is needed for affine expressions. While non-affine decomposition expressions such as $\beta = \sigma_g(\alpha)$ appear to have multi-dimensional inputs, because the function $\sigma_g$ operates element-wise, this is simply a shorthand notation for $N$ decomposition expressions, each of which is unique and unary.

\label{ex:LSTM}
We demonstrate the use of functional decomposition to over-approximate the input-output graph of an LSTM network $\F$ with a hybrid zonotope. $\F$ is trained to predict a noisy sine wave 
\begin{equation}
    \label{eqn:LSTM_time_eqn}
    f(t) = \sin\left(\frac{2\pi}{200}t\right)+\nu\:,
\end{equation} at discrete integer steps in $t$, where $\nu$ is a normally distributed random variable with mean 0 and standard distribution 0.05. $\F$ is trained on on the data for $t\in\{0,1,2,\dots,800\}$ for 100 epochs. The training data was not formatted as a timeseries of points $(t_i,\ f(t_i))$, but instead as a sequence of state-update points $(f(t_i),\ f(t_{i+1}))$ so that the network learns to predict how the state changes over time (instead of a dependence on time itself). 

\begin{figure}
    \centering
    \includegraphics[width=.75\linewidth, trim = {0 0 0 0in}, clip]{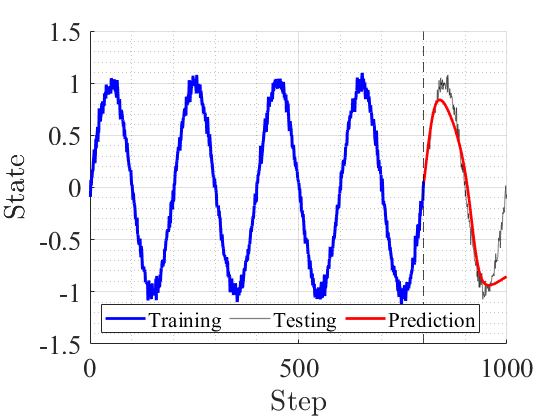}
    \caption{The LSTM network $\F$ was trained on the function $f(t)$ given in~\eqref{eqn:LSTM_time_eqn} for values $t\in\{0,1,2,\dots,800\}$, and its predictions were tested against the values for $t\in\{801,802,\dots,1000\}$.}
    \label{fig_ex_LSTM_forecast}
    \vspace{-\baselineskip}
\end{figure}

$\F$ has 5 hidden nodes, which all use a \emph{state function} of $tanh$ and a \emph{gate function} of hard-sigmoid $\bar{\sigma}$, which is defined as
\begin{equation}
    \bar{\sigma}(x) \coloneqq {\small\begin{cases}
        0 \;, & \text{if }x < -2.5 \;,\\
        0.2 x+0.5 \;, & \text{if }-2.5\leq x\leq 2.5 \;,\\
        1 \;, & \text{else}\;.
    \end{cases}}
\end{equation}
This network was chosen to be small for the sake of exposition, but as shown in Figure~\ref{fig_ex_LSTM_forecast}, $\F$ is capable of making reasonable predictions. We construct an over-approximation of the graph of the network output given an input domain of $\int{-1.14}{1.14}$ using the decomposition~\eqref{eq:lstm_decomp} and methods from~\cite{Siefert_Reach_FuncDecomp}. Techniques from~\cite{Glunt2024_ADHS} are used to generate hybrid zonotope over-approximations of the functions $\tanh$ and $(\cdot)^2$.  

\begin{figure}
    \centering
    \includegraphics[width=.75\linewidth, valign = t]{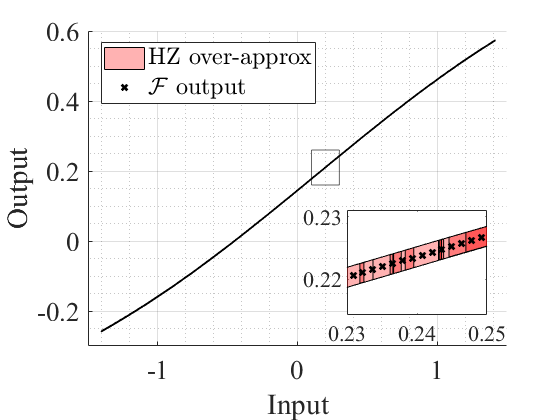}
    \caption{Hybrid zonotope (HZ) over-approximation of the network $\F$, along with sampled input-output pairs of $\F$, for the first prediction step of the network after training on the first 800 points (i.e., the hidden state and cell state have evolved to the values $h_{800}$ and $c_{800}$).}
    \label{fig:HZ_LSTM}
    \vspace{-\baselineskip}
\end{figure}

The resulting hybrid zonotope that over-approximates $\F$ is plotted in Figure~\ref{fig:HZ_LSTM}, along with points sampled in the input domain of interest. This hybrid zonotope has complexity $n_g = 8841,\ n_b = 4295,\ n_c = 4615,\ n_L = 681$, where $n_L$ is the number of ``leaves" of the hybrid zonotope (i.e., only 681 of the $2^{4295}$ combinations of binary factors result in feasible continuous factors). 

\subsection{Discrete Hybrid Automata}
\label{sect:DHAusingHZandFD}

Discrete Hybrid Automata (DHA) are a class of discrete-time hybrid systems that combine four components: An event generator (EG), a finite state machine, a mode selector (MS), and a switched affine system (SAS). The interested reader is referred to \cite[Ch. 16]{borrelli_predictive_2017} for a detailed review of DHA. 

Previous work by the authors provided reachability methods for a class of systems called Mixed-Logical Dynamical (MLD) systems \cite{Bird_HybZono,SiefertHybSUS}, and 
a DHA can be converted to an MLD system using techniques and tools such as HYSDEL~\cite{torrisi_hysdel-tool_2004}, and reachability analysis could be performed on an equivalent MLD system. This section presents an alternative approach that avoids the conversion to an equivalent MLD system by exploiting the structure of DHA subsystems via functional decomposition. The approach is demonstrated using an example DHA adopted from \cite[Ex. 16.6]{borrelli_predictive_2017}.
\allowdisplaybreaks

Consider the DHA system consisting of a continuous state $x_k\in\Rspace$, a continuous input $u_k\in\Rspace$, a mode indicator signal $i_k\in\{1,2,3\}$, and event signals $\delta_1,\delta_2\in\{0,1\}$, given by
\begin{align}
    \text{SAS: }& x_{k+1} = {\small\begin{cases}
    x_{k} + u_{k} - 1\;, & \text{if $i_k = 1$}\:,\\
    2 x_{k}\;, & \text{if $i_k = 2$}\:,\\
    2\;, & \text{if $i_k = 3$\:,}
    \end{cases}}\\
    \text{EG: }& {\small\begin{cases} 
        \delta_1 = \begin{cases}
            0\;, & \text{if } x_k < 0\:,\\
            1\;, & \text{if } x_k \geq 0\:,
        \end{cases}\\
        \delta_2 = \begin{cases}
            0\;, & \text{if } x_k + u_k -1 < 0\:,\\
            1\;, & \text{if } x_k + u_k -1 \geq 0 \:,
        \end{cases}
    \end{cases}}\\
    \text{MS: }& i_k = {\small\begin{cases}
        1\;, & \text{if } (\delta_1,\delta_2) = (0,0)\:,\\ 
        2\;, & \text{if } \delta_1 = 1\:,\\ 
        3\;, & \text{if } (\delta_1,\delta_2) = (0,1)\:.
    \end{cases}}
\end{align}
Using \textbf{Algorithms 2} and \textbf{3}, a functional decomposition is obtained as
\vspace{-11pt}
{\small\begin{align*}
        w_1 &\leftarrow x_k\;,\\
            w_2 &\leftarrow u_k\;,\\
        w_3 &= \begin{cases}
            0\;, & \text{if } w_1 < 0\:,\\
            1\;, & \text{if } w_1 \geq 0\:,
        \end{cases}\\
        w_4 &= w_1 + w_2 -1\;,\\
        w_5 &= \begin{cases}
            0\;, & \text{if } w_4 < 0\:,\\
            1\;, & \text{if } w_4 \geq 0 \:,
        \end{cases}\\
        w_6 &= \begin{cases}
        1\;, & \text{if } (w_3,w_5) = (0,0)\:,\\ 
        2\;, & \text{if } w_3 = 1\:,\\ 
        3\;, & \text{if } (w_3,w_5) = (0,1)\:,
        \end{cases}\\
        w_7 &= \begin{cases}
            1\;, & \text{if } w_6 = 1\:,\\
            0\;, & \text{if } w_6 = 2 \vee w_6 = 3\:,
        \end{cases}\\
        w_8 &= \begin{cases}
            1\;, & \text{if } w_6 = 2\:,\\
            0\;, & \text{if } w_6 = 1 \vee w_6 = 3\:,
        \end{cases}\\
        w_9 &= \begin{cases}
            1\;, & \text{if } w_6 = 3\:,\\
            0\;, & \text{if } w_6 = 1 \vee w_6 = 2\:,
        \end{cases}\\
        w_{10} &= 2 w_1\:,\\
        w_{11} &= w_7 w_{4}\:,\\
        w_{12} &= w_8 w_{10}\:,\\
        w_{13} &= w_{10} + w_{11} + 2w_{12} \leftarrow \text{Output.}
\end{align*}}%
The DHA is represented by a functional decomposition with only unary and binary functions with the exception of $w_{13}$, which is affine.  \textbf{Algorithm 2} produces a decomposition with 16 observables, but \textbf{Algorithm 3} reduces that to the 13 shown above. Using results reported in~\cite{Siefert_Reach_FuncDecomp, Siefertthesis_2024}, hybrid zonotopes can be efficiently constructed to \textit{exactly} represent each of the nonlinear functions with low memory complexities, with the exception of $w_3$ and $w_5$. 

Hybrid zonotopes can tightly inner-approximate or over-approximate the functions for $w_3$ and $w_5$ over a domain $\int{\underline{x}}{\Bar{x}}$ with $\underline{x}<0$ and $\Bar{x}>0$ using~\cite[Thm. 5]{Siefert_Reach_FuncDecomp} with the vertex and incidence matrices %

\vspace{-11pt}
{\small\begin{align}
    V&=\begin{bmatrix}
        \underline{x} & a & 0 & \Bar{x}\\
        0 & 0 & 1 & 1
    \end{bmatrix}\:,\\
    M&=\begin{bmatrix}
        1 & 1 & 0 & 0\\
        0 & 0 & 1 & 1
    \end{bmatrix}^T\:,
\end{align}}%
where $a=0$ yields an over-approximation and $\underline{x} \leq a <0$ yields an inner-approximation. To provide the largest inner approximation achievable by a computer, let $a$ be the negation of machine precision.
\section{Conclusion}
\label{sect:conclusion}
This paper proposes algorithms that automate the task of functional decomposition. The methods are tailored to use in hybrid zonotope methods for reachability analysis and verification by mitigating excessive unary decomposition and simplifying a functional decomposition to an equivalent form with fewer observables. These functional decomposition techniques are applied to construct a set that over-approximates the input-output graph of an LSTM RNN, and a new approach for creating a set-based representation of a DHA. Future work will apply functional decomposition to other neural network architectures and further explore the relationships between functional decomposition and other set representations besides hybrid zonotopes.

\bibliographystyle{IEEEtran}
\bibliography{bibNew}

\begin{thebibliography}{10}
\providecommand{\url}[1]{#1}
\csname url@samestyle\endcsname
\providecommand{\newblock}{\relax}
\providecommand{\bibinfo}[2]{#2}
\providecommand{\BIBentrySTDinterwordspacing}{\spaceskip=0pt\relax}
\providecommand{\BIBentryALTinterwordstretchfactor}{4}
\providecommand{\BIBentryALTinterwordspacing}{\spaceskip=\fontdimen2\font plus
\BIBentryALTinterwordstretchfactor\fontdimen3\font minus \fontdimen4\font\relax}
\providecommand{\BIBforeignlanguage}[2]{{%
\expandafter\ifx\csname l@#1\endcsname\relax
\typeout{** WARNING: IEEEtran.bst: No hyphenation pattern has been}%
\typeout{** loaded for the language `#1'. Using the pattern for}%
\typeout{** the default language instead.}%
\else
\language=\csname l@#1\endcsname
\fi
#2}}
\providecommand{\BIBdecl}{\relax}
\BIBdecl

\bibitem{Siefert_Reach_FuncDecomp}
J.~A. Siefert, T.~J. Bird, A.~F. Thompson, J.~J. Glunt, J.~P. Koeln, N.~Jain, and H.~C. Pangborn, ``Reachability analysis using hybrid zonotopes and functional decomposition,'' \emph{Transactions on Automatic Control}, 2025.

\bibitem{Kolmogorov1957}
A.~N. Kolmogorov, ``On the representation of continuous functions of many variables by superposition of continuous functions of one variable and addition,'' \emph{Doklady Akademii Nauk SSSR}, vol. 114, pp. 953--956, 1957.

\bibitem{Lorentz1962}
G.~G. Lorentz, ``Metric entropy, widths, and superpositions of functions,'' \emph{The American Mathematical Monthly}, vol.~69, no.~6, pp. 469--485, 1962.

\bibitem{Diaconis1984}
P.~Diaconis and M.~Shahshahani, ``On nonlinear functions of linear combinations,'' \emph{Journal on Scientific and Statistical Computing}, vol.~5, no.~1, pp. 175--191, 1984.

\bibitem{Koppen_KolmogorovDecomp}
M.~Köppen, ``On the training of a {Kolmogorov} network,'' International Conference on Artificial Neural Networks, pp. 474--479, 2002.

\bibitem{barliga2005image}
B.~Barliga, I.~Tabus, J.~Rissanen, and J.~Astola, ``Image denoising based on {Kolmogorov} structure function for a class of hierarchical image models,'' \emph{Mathematical Methods in Pattern and Image Analysis}, vol. 5916, pp. 70--79, 2005.

\bibitem{Leni_KolmogorovImage}
P.-E. Leni, Y.~D. Fougerolle, and F.~Truchetet, ``Kolmogorov superposition theorem and its application to multivariate function decompositions and image representation,'' International Conference on Signal Image Technology and Internet Based Systems, pp. 344--351, 2008.

\bibitem{HechtNielsen1987KolmogorovsMN}
R.~Hecht-Nielsen, ``Kolmogorov's mapping neural network existence theorem,'' International Conference on Neural Networks, pp. 11--13, 1987.

\bibitem{Hornik1989_UniversalApproximator}
K.~Hornik, M.~Stinchcombe, and H.~White, ``Multilayer feedforward networks are universal approximators,'' \emph{Neural Networks}, vol.~2, no.~5, pp. 359--366, 1989.

\bibitem{Cybenko1989_UniversalApproximator}
G.~Cybenko, ``Approximation by superpositions of a sigmoidal function,'' \emph{Mathematics of Control, Signals and Systems}, vol.~2, p. 303–314, 1989.

\bibitem{liu2024kankolmogorovarnoldnetworks}
Z.~Liu, Y.~Wang, S.~Vaidya, F.~Ruehle, J.~Halverson, M.~Soljačić, T.~Y. Hou, and M.~Tegmark, ``{KAN: Kolmogorov-Arnold} networks,'' arXiv 2404.19756, 2024.

\bibitem{wanufelle2007global}
E.~Wanufelle, ``A global optimization method for mixed integer nonlinear nonconvex problems related to power systems analysis,'' Facultés Universitaires Notre-Dame de la Paix, Namur, Belgium, 2007.

\bibitem{leyffer2008branch}
S.~Leyffer, A.~Sartenaer, and E.~Wanufelle, ``Branch-and-refine for mixed-integer nonconvex global optimization,'' \emph{Mathematics and Computer Science Division, Argonne National Laboratory}, vol.~39, pp. 40--78, 2008.

\bibitem{Dijkstra1961SYA}
E.~W. Dijkstra, ``Algol 60 translation: An algol 60 translator for the x1 and making a translator for algol 60,'' Mathematisch Centrum, Tech. Rep., 1961.

\bibitem{KVASNICA_2011_AutoDecomp}
M.~Kvasnica, A.~Szücs, and M.~Fikar, ``Automatic derivation of optimal piecewise affine approximations of nonlinear systems,'' IFAC World Congress, pp. 8675--8680, 2011.

\bibitem{SZUCS_2012_OptimalPWA}
A.~Szűcs, M.~Kvasnica, and M.~Fikar, ``Optimal piecewise affine approximations of nonlinear functions obtained from measurements,'' Conference on Analysis and Design of Hybrid Systems, pp. 160--165, 2012.

\bibitem{Bemporad2005}
A.~Bemporad, A.~Garulli, S.~Paoletti, and A.~Vicino, ``A bounded-error approach to piecewise affine system identification,'' \emph{Transactions on Automatic Control}, vol.~50, no.~10, pp. 1567--1580, 2005.

\bibitem{Asarin2003}
E.~Asarin, T.~Dang, and A.~Girard, ``Reachability analysis of nonlinear systems using conservative approximation,'' Hybrid Systems: Computation and Control, pp. 20--35, 2003.

\bibitem{Siefertthesis_2024}
J.~Siefert, ``Reachability analysis of nonlinear and hybrid systems using hybrid zonotopes and graphs of functions,'' The Pennsylvania State University Graduate School, 2024.

\bibitem{Glunt2024_ADHS}
J.~J. Glunt, J.~A. Siefert, A.~F. Thompson, and H.~C. Pangborn, ``Error bounds for compositions of piecewise affine approximations,'' Conference on Analysis and Design of Hybrid Systems, pp. 43--50, 2024.

\bibitem{Bird_HybZono}
T.~J. Bird, H.~C. Pangborn, N.~Jain, and J.~P. Koeln, ``Hybrid zonotopes: A new set representation for reachability analysis of mixed logical dynamical systems,'' \emph{Automatica}, vol. 154, p. 111107, 2023.

\bibitem{Gross2018_GraphTheory}
J.~Gross, J.~Yellen, and M.~Anderson, \emph{Graph Theory and Its Applications}, 3rd~ed.\hskip 1em plus 0.5em minus 0.4em\relax Chapman and Hall/CRC, 2018.

\bibitem{Wolle2004}
T.~Wolle and H.~L. Bodlaender, ``A note on edge contraction,'' Institute of Information and Computing Sciences, Utrecht University, Tech. Rep., 2004.

\bibitem{SparsePolyZono_Althoff}
N.~Kochdumper and M.~Althoff, ``Sparse polynomial zonotopes: A novel set representation for reachability analysis,'' \emph{Transactions on Automatic Control}, vol.~66, no.~9, pp. 4043--4058, 2021.

\bibitem{LSTM}
S.~Hochreiter and J.~Schmidhuber, ``{Long Short-Term Memory},'' \emph{Neural Computation}, vol.~9, no.~8, pp. 1735--1780, 11 1997.

\bibitem{Ortiz2023}
J.~Ortiz, A.~Vellucci, J.~Koeln, and J.~Ruths, ``Hybrid zonotopes exactly represent {ReLU} neural networks,'' Conference on Decision and Control, pp. 5351--5357, 2023.

\bibitem{Zhang2023}
Y.~Zhang, H.~Zhang, and X.~Xu, ``Backward reachability analysis of neural feedback systems using hybrid zonotopes,'' \emph{Control Systems Letters}, vol.~7, pp. 2779--2784, 2023.

\bibitem{akintunde2019verification}
M.~E. Akintunde, A.~Kevorchian, A.~Lomuscio, and E.~Pirovano, ``Verification of {RNN}-based neural agent-environment systems,'' Conference on Artificial Intelligence, pp. 6006--6013, 2019.

\bibitem{jacoby2020verifying}
Y.~Jacoby, C.~Barrett, and G.~Katz, ``Verifying recurrent neural networks using invariant inference,'' Automated Technology for Verification and Analysis, pp. 57--74, 2020.

\bibitem{tran2023verification}
H.~D. Tran, S.~W. Choi, X.~Yang, T.~Yamaguchi, B.~Hoxha, and D.~Prokhorov, ``Verification of recurrent neural networks with star reachability,'' International Conference on Hybrid Systems: Computation and Control, pp. 1--13, 2023.

\bibitem{Choi2025_RNN_StarSets}
S.~W. Choi, Y.~Li, X.~Yang, T.~Yamaguchi, B.~Hoxha, G.~Fainekos, D.~Prokhorov, and H.-D. Tran, ``Reachability analysis of recurrent neural networks,'' \emph{Nonlinear Analysis: Hybrid Systems}, vol.~56, p. 101581, 2025.

\bibitem{borrelli_predictive_2017}
F.~Borrelli, A.~Bemporad, and M.~Morari, ``Predictive control for linear and hybrid systems,'' Cambridge University Press, 2017.

\bibitem{SiefertHybSUS}
J.~A. Siefert, T.~J. Bird, J.~P. Koeln, N.~Jain, and H.~C. Pangborn, ``Robust successor and precursor sets of hybrid systems using hybrid zonotopes,'' \emph{Control Systems Letters}, vol.~7, pp. 355--360, 2023.

\bibitem{torrisi_hysdel-tool_2004}
F.~Torrisi and A.~Bemporad, ``{HYSDEL}-a tool for generating computational hybrid models for analysis and synthesis problems,'' \emph{Transactions on Control Systems Technology}, vol.~12, no.~2, pp. 235--249, 2004.

\end{thebibliography}

\end{document}